\documentclass[
 reprint,
 superscriptaddress,
 amsmath,amssymb,
 aps,
 floatfix]{revtex4-2}

\usepackage{graphicx}
\usepackage[colorlinks=true, allcolors=blue]{hyperref}
\usepackage[per-mode=symbol]{siunitx}
\usepackage{float}
\usepackage{braket}
\usepackage{xspace}

\usepackage{amsmath}
\usepackage{mathtools}

\DeclareSIUnit \dbc{dBc}

\usepackage{dcolumn}
\usepackage{bm}

\begin{document}

\def \system {
\begin{figure}
    \includegraphics[width=\linewidth]{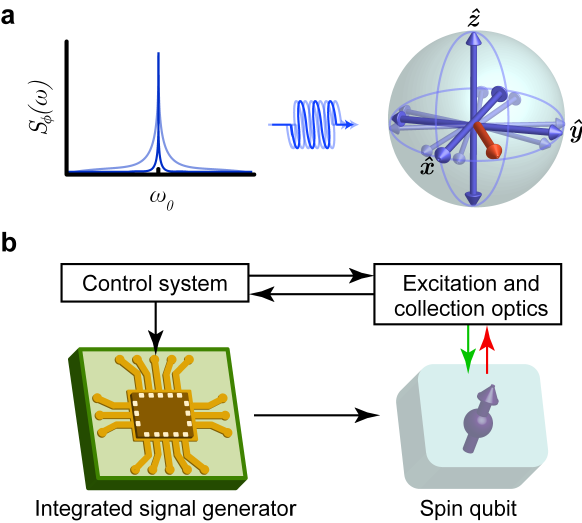}
    \caption[Effects of Oscillator Noise on a Central Spin]{\label{fig:system}\textbf{Effects of oscillator noise on a central spin.} \textbf{a}, Schematic representation of effects of phase fluctuations on axes of rotation for a central spin, where phase fluctuations (faded curve) lead to fluctuations in the rotation axes (faded axes). \textbf{b}, System diagram of a quantum measurement setup incorporating an integrated signal general for an optically addressed spin qubit.}
\end{figure}
}

\def \filterfunc {
\begin{figure}
    \includegraphics[width=\linewidth]{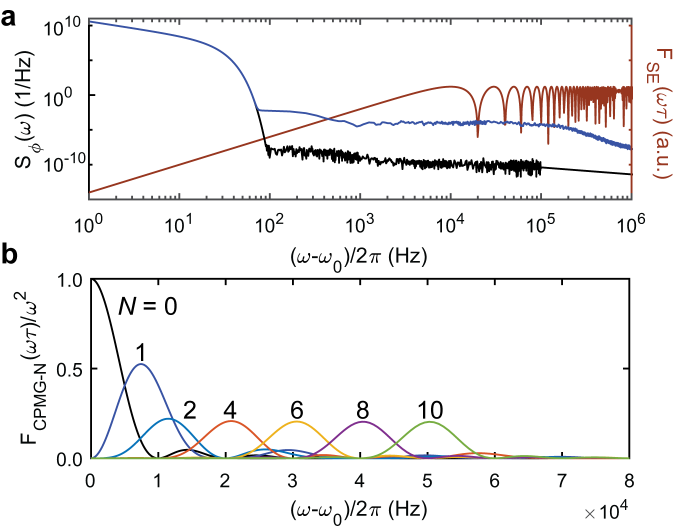}
    \caption[Control Sequences as filter-transfer functions]{\label{fig:filterfun}\textbf{Control sequences as filter-transfer functions.} \textbf{a}, filter-transfer function of a spin-echo sequence with \SI{100}{\micro\second} evolution time (maroon, right axis) plotted against the phase spectra of the $^{13}$C nuclear spin bath in conjunction with low-noise local oscillator (black, left axis) and noisy local oscillator (blue). \textbf{b}, CPMG-$N$ sequences represented as filter-transfer functions for a fixed total evolution time of \SI{100}{\micro\second}, where $N$ signifies the number of $\pi$-pulses.}
\end{figure}
}

\def \vcopll {
\begin{figure}
    \includegraphics[width=\linewidth]{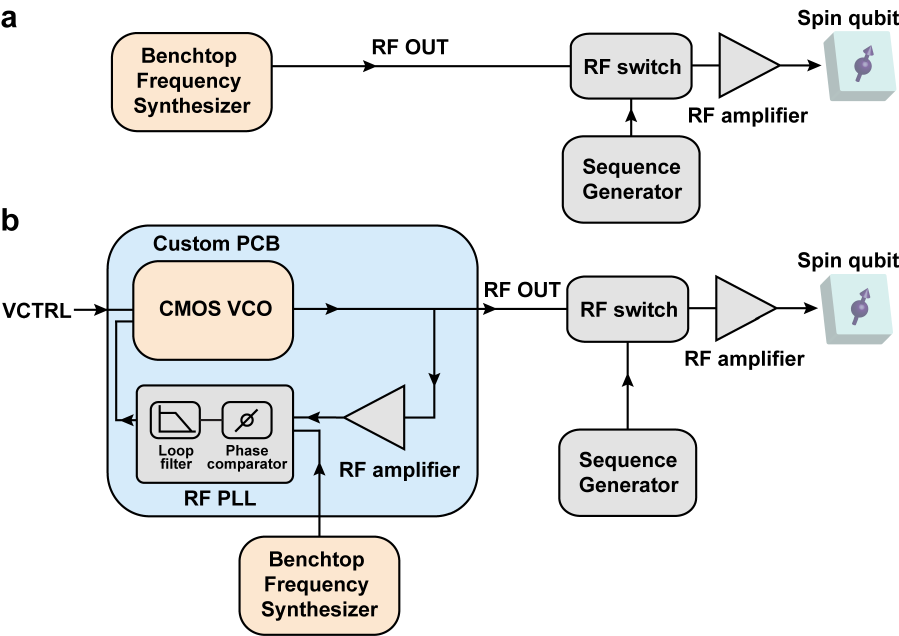}
    \caption[Schematic diagram of rf control signal delivery setup]{\label{fig:vcopll}\textbf{Schematic diagram of rf control signal delivery setup.} \textbf{a}, Experimental setup to deliver the rf control signal using a benchtop frequency synthesizer. \textbf{b}, Experimental setup to inject noise into the rf control signal, where the phase and frequency of an integrated voltage-controlled oscillator (VCO) is locked to an external frequency synthesizer through a phase-locked loop (PLL). The integrated VCO, off-chip PLL, and amplifier are interfaced through a custom printed circuit board (PCB).}
\end{figure}
}

\def \pneffects {
\begin{figure*}[t!]
    \includegraphics[width=\textwidth]{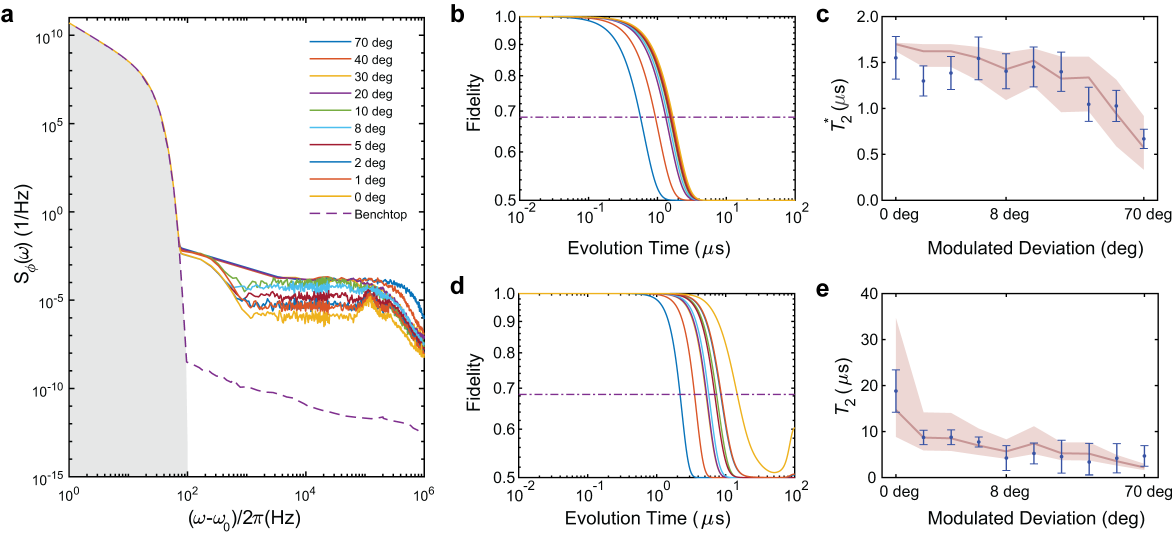}
    \caption[Effects of Phase Noise on Measured Coherence]{\label{fig:pneffects}\textbf{Effects of phase noise on measured coherence.} \textbf{a}, Measured phase spectra of noise-injected local oscillator (solid traces) and reported phase spectrum of low-noise local oscillator (purple, dashed), combined with noise spectrum of $^{13}$C nuclear spin bath (shaded grey). \textbf{b}, Free-induction decay envelopes calculated from combined noise spectra for noisy local-oscillator in (\textbf{a}). Dot-dashed line denotes the dephasing threshold. \textbf{c}, Measured (errorbars) and calculated (orange solid line) $T^\ast_2$ times, with upper and lower bounds propagated from phase noise measurement uncertainty (shaded light orange). \textbf{d}, spin-echo decay envelopes calculated from combined noise spectra for noisy local-oscillator in (\textbf{a}). Dot-dashed line denotes the dephasing threshold. \textbf{e}, Measured (errorbars) and calculated (orange solid line) spin-echo $T_2$ times, with upper and lower bounds propagated from phase noise measurement uncertainty (shaded light orange).}
\end{figure*}
}

\def \decoherence {
\begin{figure}
    \includegraphics[width=\linewidth]{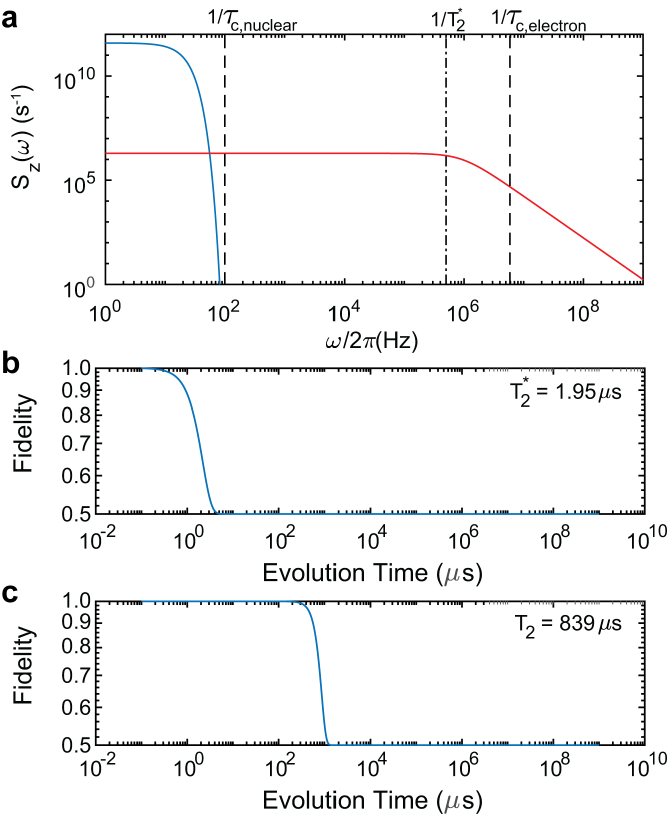}
    \caption[Decoherence from Environmental Spin Baths]{\label{fig:decoherence}\textbf{Decoherence from environmental spin baths.} \textbf{a}, Noise power spectral densities calculated for $^{13}$C nuclear spin bath (blue) and nitrogen electron spin bath (red). Dashed lines denote the respective correlation frequency for carbon nuclear spin bath and nitrogen electron spin bath. Dot-dashed line denotes the inhomogenously broadened decoherence rate measured for the studied sample. \textbf{b-c}, Decoherence envelopes calculated from $^{13}$C nuclear spin bath noise power spectral density in (\textbf{a}) for (\textbf{b}) free-induction decay and (\textbf{c}) spin-echo, with a predicted $T^\ast_2 = \SI{1.95}{\micro\second}$ and $T_2 = \SI{839}{\micro\second}$. }
\end{figure}
}

\def \jitter {
\begin{figure}[h!]
    \centering
    \includegraphics[width=\linewidth]{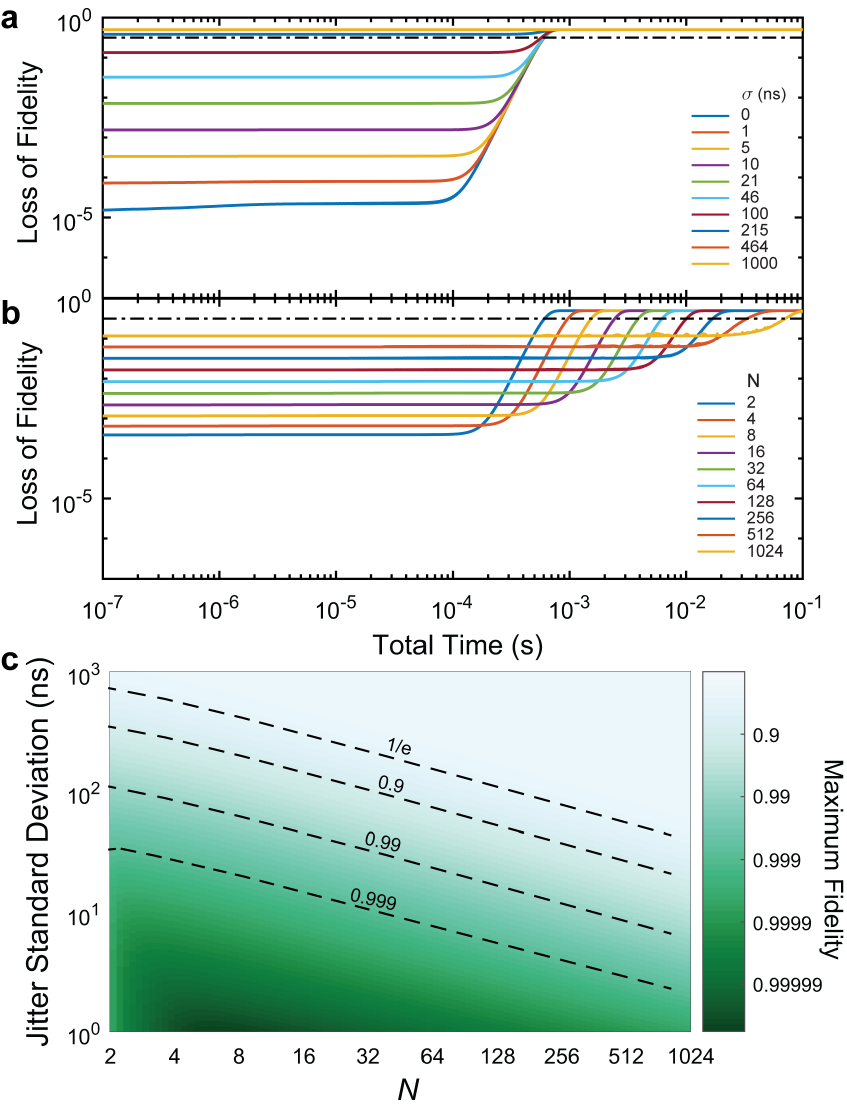}
    \caption[Contribution from Signal Jitter towards Decoherence]{\textbf{Contribution from signal jitter towards decoherence.} \textbf{a-b}, Calculated loss of fidelity from combined phase noise of $^{13}$C nuclear spin bath and low-noise local oscillator under CPMG filter function envelopes with (\textbf{a}), N = 2 and varying jitter and (\textbf{b}), varying N and jitter = \SI{5}{\nano\second}. \textbf{c}, Maximum achievable fidelity calculated from combined phase noise of $^{13}$C nuclear spin bath and low-noise local oscillator with CPMG filter function envelopes of varying $N$ and varying timing jitter.}
    \label{fig:jitter}
\end{figure}
}

\def \cpmgdeco {
\begin{figure}[t!]
    \centering
    \includegraphics[width=\linewidth]{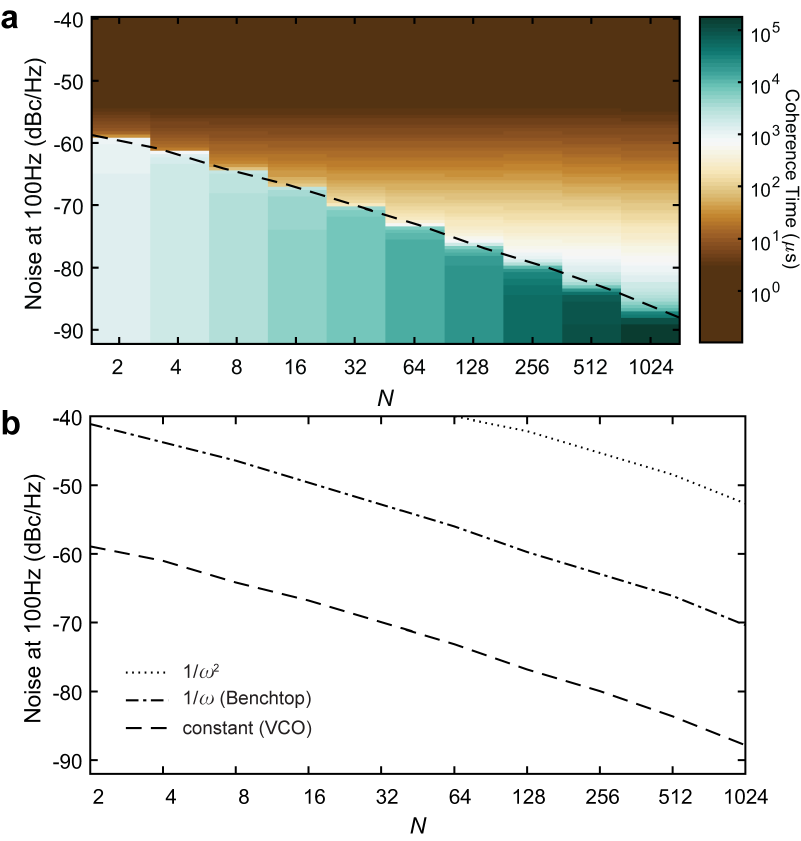}
    \caption[Oscillator-noise-limited Thresholds for CPMG Sequences]{\textbf{Oscillator-noise-limited threshold for CPMG sequences.} \textbf{a}, Coherence time calculated with CPMG sequences of varying $N$ for phase noise modeled after the integrated VCO in conjunction with a $^{13}$C nuclear spin bath. \textbf{b}, Phase noise thresholds at which the central spin coherence becomes limited by oscillator noise through CPMG sequences of varying $N$, calculated for phase noise modeled after frequency dependencies of the integrated VCO (dashed), the benchtop synthesizer (dot-dashed), and $1/\omega^2$ (dotted). The thresholds are extracted from coherence times in (\textbf{a}) calculated for each frequency dependence and is defined as the noise level at which the coherence time deteriorates past $1/e$ of the maximum coherence time.}
    \label{fig:cpmg}
\end{figure}
}


\title{
Electronic Noise Considerations for \\ Designing Integrated Solid-State Quantum Memories
}

\author{Tzu-Yung Huang}
\email[Corresponding author: ]{tzu-yung.huang@nokia-bell-labs.com}
\altaffiliation[Present Address: ]{Nokia Bell Labs, 600 Mountain Ave. Murray Hill, NJ 07974, USA}
\affiliation{%
 Quantum Engineering Laboratory, Department of Electrical and Systems Engineering, University of Pennsylvania
}%
\author{David A. Hopper}%
\altaffiliation[Present Address: ]{imec, 220 Montgomery Street, Suite 1027 San Francisco, CA 94104, USA}
\affiliation{%
 Quantum Engineering Laboratory, Department of Electrical and Systems Engineering, University of Pennsylvania
}%
\author{Kaisarbek Omirzakhov}
\affiliation{%
 Department of Electrical and Systems Engineering, University of Pennsylvania
}%
\author{Mohamad Hossein Idjadi}
\altaffiliation[Present Address: ]{Nokia Bell Labs, 600 Mountain Ave. Murray Hill, NJ 07974, USA}
\affiliation{%
 Department of Electrical and Systems Engineering, University of Pennsylvania
}%
\author{S. Alexander Breitweiser}%
\affiliation{%
 Quantum Engineering Laboratory, Department of Electrical and Systems Engineering, University of Pennsylvania
}%
\affiliation{Department of Physics, University of Pennsylvania}
\author{Firooz Aflatouni}
\affiliation{%
 Department of Electrical and Systems Engineering, University of Pennsylvania
}%
\author{Lee C. Bassett}
\email[Corresponding author: ]{lbassett@seas.upenn.edu}
\affiliation{%
 Quantum Engineering Laboratory, Department of Electrical and Systems Engineering, University of Pennsylvania
}%

\date{\today}

\begin{abstract}
As quantum networks expand and are deployed outside research laboratories, a need arises to design and integrate compact control electronics for each memory node.
It is essential to understand the performance requirements for such systems, especially concerning tolerable levels of noise, since these specifications dramatically affect a system's design complexity and cost.
Here, using an approach that can be easily generalized across quantum-hardware platforms, we present a case study based on nitrogen-vacancy (NV) centers in diamond.
We model and experimentally verify the effects of phase noise and timing jitter in the control system in conjunction with the spin qubit's environmental noise.
We further consider the impact of different phase noise characteristics on the fidelity of dynamical decoupling sequences.
The results demonstrate a procedure to specify design requirements for integrated quantum control signal generators for solid-state spin qubits, depending on their coherence time, intrinsic noise spectrum, and required fidelity.
\end{abstract}

\maketitle

\section{Introduction}
The realization of quantum networks, where information from separate, independent quantum computing units or sensors can be relayed and stored, is crucial to the development of distributed quantum computing \cite{Cuomo2020, Gill2022}, quantum communication networks \cite{Cacciapuoti2020}, and the practical deployment of quantum technologies \cite{Guo2019, Zhang2021}. 
In particular, quantum memories serve as the backbone of such a network by allowing quantum information to be stored locally and distributed with other connected nodes \cite{Cho2016, Drmota2023}. 
Among the promising platforms for implementing quantum memories are optically interfaced solid-state spins \cite{Awschalom2018}. 
These platforms enable the configuration of a hybrid, multi-qubit quantum register formed by an electron spin coupled to nearby nuclear spins. 
With built-in access to long-lived nuclear spins that serve as memory nodes, the optically addressable electron spins preserve their accessibility to the outside environment with a spin-photon interface to faciliate sensing, computation, and communication \cite{Robledo2011,Maurer2012,Degen2017,Sukachev2017,Kalb2018}. 
The technology and performance of such memory-enhanced quantum networks are rapidly advancing \cite{Bhaskar2020, Bradley2022, Pompili2021}. 
\system

As the size of the quantum network increases, device performance and complexity of each node are critical considerations for network scalability and stability. 
While much effort has gone into integrating optical interfaces \cite{Huang2019, Wan2020, Palm2023}, equally important are the electronics responsible for signal generation and conditioning. 
For solid-state spin qubit systems, the ability to accurately output microwave control signals of the desired frequencies and pulse shapes is paramount to their capabilities and performance as quantum information nodes. 
Since the frequency and phase of the local oscillator set the axes of rotation to coherently drive the central spin, frequency drift and phase noise can lead to cumulative dephasing of the quantum system (Fig.~\ref{fig:system}\textbf{a}) \cite{Ball2016}. 
Typically, optimized benchtop electronics are used for coherently controlling solid-state spin qubits due to their large bandwidth, low noise levels, and high stability in frequency and phase. 
However, benchtop systems are inhibitive to scaling and integrating quantum memories into a practical quantum network, especially as the number of network nodes increases. 
In most cases, much of the benchtop system's bandwidth is unused and its specifications could be optimized to maximize efficiency. 
There is potential for compact, integrated control signal sources to be leveraged here for their low power consumption, small footprint, and reconfigurability (Fig.~\ref{fig:system}\textbf{b}) \cite{Omirzakhov2023}. 
However, to successfully optimize for each quantum system, it is crucial to understand the design parameters of an integrated control signal source and how they impact the coherence of the interfaced qubit. 

Here, we use the filter-transfer function formalism to examine the impact of local oscillator noise on a central spin in conjunction with its environmental noise. 
The filter-transfer function formalism is commonly used to analyze effects of environmental noise, as well as its mitigation through advanced control sequences, on a central spin \cite{Uhrig2007, Uhrig2008, Ball2016, Yang2017}. 
We employ the nitrogen-vacancy (NV) center in diamond, a leading spin qubit platform for quantum memories, as a case study. 
For coherence-limited applications such as quantum memories and registers, dynamical decoupling sequences are typically employed to mitigate dephasing induced by the central NV electron spin's local environment and to individually address nearby nuclear spins \cite{Taminiau2012, Bradley2019}. 
The number of control pulses can exceed $10^3$, at which point coherence often becomes limited by noise in the control electronics \cite{Bar-Gill2013}. 
In this work, we examine the impact of classical control noise on the coherence of a central NV electron spin in a naturally occurring \SI{1.1}{\%} $^{13}$C nuclear spin bath through dynamical decoupling sequences. 
We measure and verify this analytical framework with coherent measurements of a single NV center in bulk diamond. 
The framework is easily generalizable to other quantum hardware platforms, as a means to define the performance requirements for classical control electronics.

\section{Decoherence from Environmental Spin Baths}

Before discussing effects of noise arising from an external oscillator, we first consider the effects of environmental noise intrinsic to a spin-qubit's host crystal. 
Paramagnetic defects and impurities distributed throughout the host material give rise to fluctuating magnetic fields. 
In synthetic diamond, the dominant noise sources are typically electron spins from substitutional nitrogen atoms and nuclear spins associated with $^{13}$C isotopes \cite{Maze2008,DeLange2010,Reinhard2012,Zhao2012,Wang2013}. 
For typical ``optical grade'' chemical vapor deposition (CVD) grown diamond with $>\SI{10}{ppm}$ substitutional single-nitrogen impurities, spin coherence of NV centers is limited by interaction with N electron spin baths \cite{Wang2013}. 
On the other hand, in ultrapure ``electronic grade'' diamond where the nitrogen impurities are $<\SI{5}{ppb}$, the NV electron spin coherence is limited by hyperfine interactions with the natural \SI{1.1}{\percent} abundance of $^{13}$C isotopes in the diamond crystal \cite{Maze2008}.

We consider the effects on a central NV electron spin from nitrogen electron spin baths and carbon nuclear spin baths separately (Fig.~\ref{fig:decoherence}\textbf{a}). 
For a spin bath with large density of nitrogen electron spins and in the absence of nearby, strongly coupled electron spins, the NV center electron spin is coupled to the electron spin bath through magnetic dipole interactions \cite{Yang2017}. 
In this scenario, the electron spin bath's collective influence on the central NV electron spin is much stronger than the central electron spin's action on the spin bath. 
Accordingly, the central spin's decoherence timescale is much faster than the spin bath's evolution caused by the central spin. 
Therefore, within the qubit coherence time, the environmental noise from the electron spin bath can be approximated as classical Gaussian noise \cite{Hanson2008,Dobrovitski2009,Witzel2012,DeLange2010,Wang2013}. 

\decoherence
For a spin system impacted by Gaussian phase noise with random variable $\varphi$, the central spin dephasing after time $\tau$ is determined by the phase variance, $\langle\varphi^2(\tau)\rangle$  \cite{Yang2017}:
\begin{equation}
\label{eqn:noisecorr}
    L(\tau) = e^{-\braket{\varphi^2(\tau)}/2} = e^{-\chi(\tau)},
\end{equation}
\noindent where $\chi(\tau)$ denotes the decay envelope \cite{Ball2016,Bauch2020},
\begin{equation}
\label{eqn:noiseintegral}
    \chi(\tau) = \frac{1}{\pi} \int_{0}^{\infty} \frac{d\omega}{\omega^2} S_z(\omega)F(\omega\tau).
\end{equation}
Here, $F(\omega\tau)$ is the square modulus of the frequency-domain control vector \cite{Green2013},
\begin{equation}
    \label{eqn:filterdef}
    F(\omega\tau) = |y(\omega\tau)|^2,
\end{equation}
\noindent and $S_z(\omega)$ is the dephasing noise power spectral density (PSD). 
In other words, $\chi(\tau)$ expresses the overlap integral between the dephasing noise PSD and the control-sequence filter-transfer function. 
The dephasing noise PSD of an electron spin bath is typically modeled by the Ornstein-Uhlenbeck noise \cite{Bauch2020,Dobrovitski2009,Witzel2012,DeLange2010},
\begin{equation}
\label{eqn:PSDdef}
    S_{z,\mathrm{electron}}(\omega) = \frac{\Delta^2\tau_c}{\pi} \frac{1}{1+(\omega\tau_c)^2},
\end{equation}
\noindent where $\Delta$ corresponds to the average coupling strength between the central NV electron spin and the nitrogen electron bath spins, and $\tau_c$ is the correlation time of the nitrogen electron bath and is related to their characteristic flip-flop time \cite{Bar-Gill2012}. 

For a characteristic dephasing PSD of the nitrogen electron spin bath, we use the values calculated for a typical optical-grade, CVD grown diamond with \SI{100}{ppm} substitutional-nitrogen density by Bar-Gill \textit{et al}. \cite{Bar-Gill2012}
The resulting $S_{z,\mathrm{electron}}(\omega)$ calculated with Eqn.~(\ref{eqn:PSDdef}) is shown in Fig.~\ref{fig:decoherence}\textbf{a} (red curve). 
The noise spectrum has a soft high-frequency cutoff defined by $1/\tau_\mathrm{c,electron}$. 
When $\tau_\mathrm{c} \ll \tau$, as is the case for NV centers in nitrogen electron spin baths, correlations within the environment decay before evolution caused by the central NV electron spin can be accrued, meaning that the back-action from the central NV electron spin can be ignored.

However, this is not the case of an NV-center whose decoherence is dominated by naturally occurring \SI{1.1}{\percent} $^{13}$C nuclear spins. 
In quantum memory applications, $\tau$ can often approach the $^{13}$C nuclear spin-spin correlation time of $\sim\SI{10}{\milli\second}$, meaning that the back-action from the central NV cannot necessarily be ignored \cite{Hall2014}. 
Nevertheless, the NV-center electron spin has been shown, both theoretically and experimentally, to exhibit Gaussian coherence decay when the total evolution time $\tau$ approaches the coherence time $T_2$ \cite{Zhao2012, Hall2014}. 
Furthermore, it has been demonstrated that the $^{13}$C nuclear spin bath can be approximated as Gaussian noise when operating in magnetic field $>$ 100 G, which is typically the case in quantum memory applications \cite{Reinhard2012}. 
As a result, the decay function established earlier in Eqn.~(\ref{eqn:noisecorr}) is still applicable. 
The power spectral density of the carbon nuclear spin bath, $S_{z,\mathrm{nuclear}}(\omega)$, exhibits a sharp high-frequency cutoff at $1/\tau_\mathrm{c,nuclear}$ (Fig.~\ref{fig:decoherence}\textbf{a}, blue curve) \cite{Yang2017,DeLange2010,Reinhard2012,Zhao2012}, with the form, 
\begin{equation}
    \label{eqn:PSDdefN}
    S_{z,\mathrm{nuclear}}(\omega) = \Delta^2\tau_ce^{-(\omega\tau_c)^2}.
\end{equation}

In this study, we use an ultrapure single-crystal diamond with $<\SI{5}{ppb}$ nitrogen impurity concentration (Element6, ELSC). 
Hence, the central NV electron spin in this case should be limited by $^{13}$C nuclear spins, with a noise PSD given by $S_{z,\mathrm{nuclear}}(\omega)$.
Using this model for the intrinsic decoherence, we calculate the state fidelity, $\mathcal{F}(\tau)$, of the central NV electron spin due to the nuclear spin bath under free-induction decay and spin-echo sequences using $\chi(\tau)$ from Eqn.~(\ref{eqn:noiseintegral}) \cite{Green2013},
\begin{equation}
\label{eqn:fidelity}
    \mathcal{F}(\tau) \approx \frac{1}{2}[1+e^{-\chi(\tau)}].
\end{equation}
\noindent The filter-transfer function forms of the free-induction decay (FID, Ramsey) and spin-echo are as follows \cite{Ball2016},
\begin{align}
    F_\mathrm{FID}(\omega\tau) &= 4\sin^2(\frac{\omega\tau}{2})\label{eqn:fidfilter} \\
    F_\mathrm{SE}(\omega\tau) &= 16\sin^4(\frac{\omega\tau}{4})\label{eqn:sefilter}.
\end{align}
\noindent Using Eqns.~(\ref{eqn:noiseintegral}) \& (\ref{eqn:fidelity}), the free-induction and spin-echo decay envelopes are calculated, yielding an inhomogeneous dephasing timescale, $T^\ast_2 = \SI{1.95}{\micro\second}$, and a spin-echo coherence time, $T_2 = \SI{839.31}{\micro\second}$ (Fig~\ref{fig:decoherence}\textbf{b, c}). 
The calculations are in excellent agreement with the measured $T^\ast_2 = 1.92\pm\SI{0.08}{\micro\second}$ and $T_2 = 830\pm\SI{28}{\micro\second}$ (see Supplementary Information). 

\section{The Role of Classical Control Noise}

To examine the effects of noise from the classical control signals, we study two manifestations of control error: signal timing jitter and phase noise. 
Timing jitter results from fluctuations in the arrival times of the control pulses, whereas phase noise results from fluctuations in the phase of the oscillating control signal.
In this section, we first consider the role of phase noise before incorporating timing jitter.

As depicted in Fig.~\ref{fig:system}\textbf{a}, oscillator phase noise manifests analogously to environmental dephasing noise, since fluctuations in the axes defining the qubit's rotating frame are equivalent to fluctuations in an external magnetic field. 
The phase spectrum, $S_\phi(\omega)$, is related to the dephasing noise PSD as follows \cite{Ball2016},
\begin{equation}
\label{eqn:phasePSD}
    S_\phi(\omega) = \frac{4}{\omega^2}S_z(\omega).
\end{equation}
\noindent Using Eqn.~(\ref{eqn:noiseintegral}), we can write the dephasing envelope as
\begin{equation}
\label{eqn:phasechi}
    \chi(\tau) = \frac{1}{4\pi} \int_{0}^{\infty} d\omega S_\phi(\omega) F(\omega\tau).
\end{equation}
We can further combine the effects of local oscillator phase noise with environmental noise sources by calculating the combined phase spectrum,
\begin{equation}
\label{eqn:totalPSD}
    S_\phi(\omega) = S_{\phi,\mathrm{nuclear}}(\omega) 
    + S_{\phi,\mathrm{LO}}(\omega).
\end{equation}

In the ideal scenario, fluctuations in the classical control signal is minimized such that errors are dominated by environmental sources.
This is typically the case when using benchtop electronics. 
As an example, we consider a low-noise benchtop frequency synthesizer (Stanford Research Systems, SG384) as the local oscillator driving our qubit rotations and calculate the effects of its phase noise on the qubit fidelity from Eqn.~(\ref{eqn:fidelity}) using the phase spectrum reported by the manufacturer (see Supplementary Information, Fig.~S1). 
Indeed, we find that phase noise makes a negligible contribution to free-induction decay and spin-echo control sequences; the dephasing and decoherence times presented in Fig.~\ref{fig:decoherence}\textbf{b,c} remain unchanged. 
In fact, a calculation of $\mathcal{F}(\tau)$ using only $S_{\phi,\mathrm{LO}}(\omega)$ shows that there is less than 0.01\% loss of coherence up to \SI{3}{\second} total evolution time (see Supplementary Information, Fig.~S1\textbf{b,c}).

\filterfunc
Figure~\ref{fig:filterfun}\textbf{a} illustrates the effects of phase noise and its filtering, showing $S_\phi(\omega)$ for representative cases of low- and high-noise local oscillators, along with $F_\mathrm{SE}(\omega\tau)$ calculated with a total evolution time of \SI{100}{\micro\second}. 
The phase spectra consist of the sum of the predicted environmental spin bath added to measured phase spectra for the SG384 benchtop synthesizer (black curve) and for a variable phase-noise oscillator integrated on a CMOS chip (blue curve), which will be discussed in detail later. 
In both cases, the nuclear spin bath dominates noise contributions at low frequencies, and the local oscillator takes over above the frequency cutoff for the nuclear spin bath, $1/\tau_\mathrm{c,nuclear}$. 
While the spin-echo sequence filters out most of the low-frequency noise from the nuclear spin bath, most of the high-frequency noise is passed through. 

To illustrate how such frequency filtering can be engineered, we also consider the Carr-Purcell-Meiboom-Gill (CPMG) multi-pulse control sequence. 
The Fourier transform of its modulation function with $N$ $\pi$ pulses has the form \cite{Uhrig2008}
\begin{equation}
    \label{eqn:cpmgfilter}
    y_N(\omega\tau) = 1 + (-1)^{N+1}e^{i\omega\tau} + 2\sum_{j=1}^{N}(-1)^je^{i\omega\tau_j},
\end{equation}
\noindent where $\tau$ is the total evolution time up to $N$ pulses and $\tau_j$ is the timing of the $j$th $\pi$ pulse.
Following Eqn.~(\ref{eqn:filterdef}), the filter-transfer function of CPMG sequences with even number of $\pi$-pulses is \cite{Cywinski2008}
\begin{equation}
    \label{eqn:cpmgeven}
    F_\mathrm{even}(\omega\tau) = \frac{8\mathrm{sin}^2(\frac{\omega\tau}{2})\mathrm{sin}^4(\frac{\omega\tau}{4N})}{\mathrm{cos}(\frac{\omega\tau}{2N})}
\end{equation}
\vcopll
\pneffects
Figure~\ref{fig:filterfun}\textbf{b} shows the CPMG filter-transfer function with $N=1$ to 10 for a fixed total evolution time of \SI{100}{\micro\second} on a linear scale.
Here, $N=0$ corresponds to free-induction decay and $N=1$ corresponds to spin-echo. 
The filtering effects are clarified by plotting the filter functions as $F(\omega\tau)/\omega^2$ to emphasize the parts of $F(\omega\tau)$ that produce the largest contributions to dephasing error \cite{Biercuk2011}. 
As more and more $\pi$-pulses are applied, the filter envelope shifts towards higher frequencies. 
Typically, this leads to longer coherence times since $S_\phi(\omega)$ decreases with $\omega$.

\section{Experimental results}

In order to experimentally verify the theoretical framework accounting for local oscillator phase noise on the coherence of a central spin, we employ a variable phase-noise frequency synthesizer with an integrated voltage-controlled oscillator (VCO) and an off-chip phase-locked loop (PLL). 
The experimental setup diagram is shown in Fig.~\ref{fig:vcopll}, and more circuit details are provided in supporting information; see Fig.~S4. 
The differential output of the VCO is coupled to an rf amplifier (Texas Instruments, LMH5401) before being fed to the off-chip PLL (Texas Instruments, LMX2491), which uses a frequency reference generated by a high-precision rf signal generator. 
The output of the rf PLL is connected in feedback to the voltage bias node of the on-chip VCO. 
This feedback mechanism locks both phase and frequency of the on-chip VCO to an external rf signal through the PLL. 
The rf signal is then delivered to the spin qubit through an amplifier (Mini-Circuits, ZHL-16W-43-S+) and an rf switch (Mini-Circuits, ZASWA-2-50-DR) modulated by an arbitrary waveform generator (Tektronix, AWG7102).

To inject noise to the VCO, we lock the integrated VCO to the benchtop frequency synthesizer (Stanford Research Systems, SG384) through the PLL, and then phase modulate the reference with noise. 
The applied phase noise is set with a modulation bandwidth of 1kHz at deviations of 0, 1, 2, 5, 8, 10, 20, 30, 40, and 70 degrees. 
To quantify the effects of the injected noise, the phase spectra of the VCO output are measured by normalizing the power spectrum, obtained using a spectrum analyzer (HP, 8563E), with respect to the carrier signal. 
The single-sideband phase noise up to 1 MHz frequency offset from the carrier signals are measured with a resolution of 300 Hz. 
The measured phase spectra, added to the predicted noise spectrum of the nuclear spin bath, are shown in Fig.~\ref{fig:pneffects}\textbf{a} (solid lines). 
The total phase spectrum for the reference source without phase modulation is shown as a dashed, purple trace; this is the same source as the low-noise local oscillator plotted in Fig.~\ref{fig:filterfun}\textbf{a}.

For each of the noise settings shown in Fig.~\ref{fig:pneffects}\textbf{a}, we use the VCO to drive free-induction decay and spin-echo pulse sequences on a single NV center in bulk diamond at \SI{2.1}{\giga\hertz}, and we use the phase noise measurements to calculate the expected dephasing and decoherence times. 
Figure \ref{fig:pneffects}\textbf{b} shows the predicted free-induction decay envelopes, calculated using Eqns.~(\ref{eqn:noiseintegral}), (\ref{eqn:fidfilter}), \& (\ref{eqn:totalPSD}), from which we extract the predicted $T^\ast_2$ times shown in Fig.~\ref{fig:pneffects}\textbf{c} (orange trace). 
Uncertainties in the predicted $T_2^\ast$ values (shaded orange) reflect the propagation of a $\pm \SI{5}{dBc/Hz}$ constant offset of the total phase spectra.
The predictions account for the phase noise within the measurement bandwidth only, since these frequencies dominate the decoherence effects in this experimental regime; extrapolation of the phase noise spectrum to higher frequencies shifts the predictions by amounts smaller than the prediction uncertainties.
Experimental values of $T_2^\ast$ are extracted from a fit to the free-induction decay data (see Supplemental Information).
The best-fit $T^\ast_2$ values are shown in Fig.~\ref{fig:pneffects}\textbf{c} as blue points, with error bars denoting \SI{95}{\percent} confidence intervals. 

The spin-echo decay envelopes and ${T}_2$ predictions shown in Figs.~\ref{fig:pneffects}\textbf{d} and \ref{fig:pneffects}\textbf{e} are similarly calculated for each noise level using Eqns.~(\ref{eqn:noiseintegral}), (\ref{eqn:sefilter}), \& (\ref{eqn:totalPSD}). 
Experimental values of $T_2$ are extracted from a fit to data (see Supplemental Information).
The best-fit $T_2$ values are shown as blue points in Fig.~\ref{fig:pneffects}\textbf{e}, with error bars denoting the \SI{95}{\percent} confidence interval. 

For both free-induction and spin-echo decay measurements, the experimental values of $T_2^\ast$ and $T_2$ agree with the predictions calculated from measured phase spectra.
This verifies the framework for modeling local environmental noise and predicting experimental coherence properties.

\section{Design requirements for quantum memories}

Using the experimentally verified framework, we now consider performance benchmarks for local oscillators in quantum memory applications. 
For spin qubits such as the NV center with access to nearby coupled nuclear spins, dynamical decoupling sequences are employed to selectively couple and decouple from these longer-lived qubits, forming the basis of operation for nuclear-register quantum memories. 
These sequences rely on trains of precisely timed pulses to perform conditional rotations of the coupled electron and nuclear spins. 
Additionally, they serve to decouple the electronic spin from nuclear spins not resonant with the pulse delay frequency, effectively filtering out the environmental bath. 
Dynamical decoupling sequences designed for different operations and couplings can involve more than $10^3$ individual pulses.
Hence, it is important to consider both the effects of local-oscillator phase noise as well as jitter in pulse arrival times, which can lead to significant error accumulation for high pulse numbers.

\subsection{Local-oscillator phase noise}
\cpmgdeco
Figure \ref{fig:cpmg} illustrates the effects of phase noise for CPMG sequences of varying $N$, where the frequency dependence of the phase spectrum plays a crucial role.
For example, the measured spectrum of the integrated VCO without noise injection has effectively no frequency dependence between \SI{10}{\hertz} and \SI{100}{\kilo\hertz} (see Supplemental Information Fig.~S5\textbf{b}), whereas the benchtop synthesizer phase spectrum (Fig.~S5\textbf{a}) varies significantly within this range.
In general, a synthesizer's phase spectrum will further depend on the carrier frequency as well as other factors.
To account for these variations, we model the phase spectrum using an empirical form \cite{IEEEStandard1139-20082008,McNeill2017}:

\begin{equation}
    S_\phi(\omega) = \sum_{\alpha=-2}^{+2}\frac{h_\alpha \omega_0^2\omega^\alpha}{\omega^2}
    + \frac{h_L}{1+(\frac{\omega}{\omega_L})^2},
    \label{eqn:modelpn}
\end{equation}

\noindent where $\alpha$ labels the order of the corresponding power-law term in $S_\phi$, $\omega_0$ is the carrier frequency, and $h_\alpha$ is the numerical coefficient for each noise process. The last term accounts for the effect of a PLL, which effectively acts as a low-pass filter on the phase spectrum, and $\omega_L$ designates the low-pass cutoff frequency. 

To account for different noise amplitude levels and the local spin bath, we apply a scaling factor to the modeled $S_\phi(\omega)$ before adding the $^{13}$C nuclear spin bath spectrum (see Supplemental Information). 
Since the $^{13}$C nuclear spin bath dominates the noise spectrum at low frequency regimes, the oscillator phase noise is relevant for frequencies beyond the $^{13}$C nuclear spin bath frequency cutoff, $1/\tau_{\mathrm{c,nuclear}} \approx \SI{100}{\hertz}$. 
Hence, we characterize each modeled phase spectrum by its amplitude at $f = \SI{100}{\hertz}$. 

Figure~\ref{fig:cpmg}\textbf{a} shows the predicted coherence time, calculated using the empirical phase spectrum of the VCO as a function of the noise amplitude for different values of $N$.
We define the coherence time as the total evolution time at which the fidelity decays by a factor of $1/e$.
When the local oscillator noise amplitude is small, the coherence time is dominated by the environment, and it can be extended by using CPMG sequences with increasing $N$.
When the local oscillator noise is high, the decoherence is dominated by the oscillator and independent of $N$.
In between these extremes, there exists an inflection curve that depends on $N$ and the phase noise amplitude, separating the regimes in which the decoherence is dominated by the bath and by the oscillator noise, respectively.
We define that inflection curve as the noise amplitude for each $N$ where the coherence time deteriorates past $1/e$ of the maximum coherence time; it is shown as a dashed curve in Fig.~\ref{fig:cpmg}\textbf{a}.

To illustrate the effect of different phase spectra, Fig.~\ref{fig:cpmg}\textbf{b} shows the inflection curves calculated for three different empirical forms of $S_\phi(\omega)$.
For these calculations, we use a carrier frequency of $\SI{2}{\giga\hertz}$.
The dashed curve represents results calculated from an empirical fit to the VCO phase spectrum, which has an $1/\omega^4$ ($\alpha = -2)$ dependence, but is flat between \SI{100}{\hertz}, where the environmental noise drops off, and \SI{100}{\kilo\hertz}, at the PLL filter edge.
The dot-dashed curve is calculated using an empirical fit to the benchtop synthesizer phase spectrum, which has an $1/\omega$ ($\alpha = 1)$ dependence and much smaller filtering effects.
Finally, the dotted curve represents results calculated from a phase spectrum with only a $1/\omega^2$ ($\alpha = 0$) dependence.
In each case, we follow the procedure described above and adjust the noise scaling factor to determine the same noise amplitude at $f = \SI{100}{\hertz}$.
The empirical spectra as well as the fitting coefficients are shown in the Supplemental Information.

It is evident that the oscillator phase noise requirements vary significantly as a function of $N$ and as a function of the phase spectrum.
This analysis does not account for pulse lengths.
Biercuk \textit{et al.} \cite{Biercuk2011} consider the effect of pulse lengths as a function of the ratio $\tau_\pi/\tau$, where $\tau_\pi$ denotes the $\pi$ pulse length. 
The regimes studied in this work have a ratio of $\tau_\pi/\tau < 10^{-2}$. 
As predicted, calculations that include pulse-length corrections have a negligible effect on the predictions of Fig.~\ref{fig:cpmg}.
We also do not account for the dc magnetic field strength, which affects the system's coherence time, or $T_1$ relaxation, which will eventually limit the coherence time independent of bath and oscillator noise.
However, with knowledge of the predicted spectra of environmental noise and local oscillator noise, we can evaluate the local oscillator's suitability for a targeted application.

\subsection{Timing jitter}
\jitter
To evaluate the effects of timing jitter, we take the CPMG Fourier transform from Eqn.~(\ref{eqn:cpmgfilter})  and make the assumption that each pulse arrival has independent deviation, $\tilde\tau_j = \tau_j + \delta_j$,
\begin{align}
    \label{eqn:jitterft}
    &y_\mathrm{N-jitter}(\omega\tilde{\tau}) = \nonumber \\
    &e^{-i\omega\delta_0} + (-1)^{N+1}e^{i\omega\tilde{\tau}_{N+1}} + 2\sum_{j=1}^{N}(-1)^je^{i\omega\tilde{\tau_j}},
\end{align}
\noindent where $\tilde{\tau_j}$ designates the delay between each pulse arrival. If we assume a Gaussian distribution for $\delta_j$, computing $|y_\mathrm{n-jitter}(\omega\tilde{\tau})|^2$ gives a modified filter function:
\begin{equation}
    \label{eqn:jitterfilter}
    \tilde{F}(\omega\tau) = (2N+2)(1-e^{\sigma^2\omega^2}) + e^{\sigma^2\omega^2} F(\omega\tau),
\end{equation}
\noindent where $N$ indicates the number of $\pi$-pulses and $\sigma^2$ is the variance. For details on the derivation, see Supplementary Information Eqn.~(S4). 

Using $F_\mathrm{even}(\omega\tau)$ from Eqn.~(\ref{eqn:cpmgeven}), we compute $\tilde{F}(\omega\tau)$ with varying amounts of timing jitter and $N$, and we calculate the loss of fidelity assuming the phase spectrum from a low-noise oscillator  (Fig.~\ref{fig:filterfun}\textbf{a}, black trace). 
The results are shown in Fig.~\ref{fig:jitter}. 
For a fixed $N$, increasing the amount of jitter introduces a fidelity ceiling | an upper-bound on the highest achievable fidelity (Fig.~\ref{fig:jitter}\textbf{a}). 
On the other hand, while increasing the number of $\pi$-pulses extends the coherence time of the central spin, it exacerbates this bounding effect on the maximum achievable fidelity (Fig.~\ref{fig:jitter}\textbf{b}). 
Intuitively, increased $N$ leads to a larger overall effect from accumulated pulse-timing errors. 
Figure~\ref{fig:jitter}\textbf{c} illustrates this effect through a contour map of the maximum achievable fidelity as a function of jitter standard deviation and $N$. 
Assuming uncorrelated, Gaussian-distributed jitter in pulse timing, a jitter standard deviation less than \SI{3}{\nano\second} is needed to achieve a fidelity of 0.999 for a CPMG sequence with $N$ = 1024.

\section{Discussion}
Since device footprint, complexity, and power consumption are important considerations for integration, it is essential to define design parameters tailored for the chosen quantum platform in the targeted applications. 
Specifying the appropriate noise requirements without over-constraining is essential for guiding design, especially considering the inherent trade-off between frequency bandwidth and phase stability. 
Similarly, minimizing timing jitter is important for achieving high fidelity operations in general, but timing jitter can be deprioritized for applications where fewer pulses or shorter pulses are required. 
Our analysis shows that, at least in this experimental instance, phase stability is a more dominant limitation than timing jitter.

This approach can be extended to account for additional effects and complexities present in real-world systems.
We do not consider amplitude fluctuations of the local oscillator, amplifier, or other circuit elements.
Amplitude noise leads to pulse infidelities and loss of coherence for long multi-pulse sequences.
We also assume a fixed $1/\omega^2$ relationship between $S_\phi(\omega)$ and $S_z(\omega)$ in Eqn.~(\ref{eqn:phasePSD}).
While the phase noise spectra shown in Fig.~\ref{fig:pneffects} account for the dominant effects observed in our experiments, higher-frequency components will play a role in other regimes. 
Ideally, the phase spectrum should be captured up to the regime where the white noise of measurement electronics dominates the signal.
With regards to timing jitter, we assume uncorrelated, Gaussian errors in each timing event, whereas noise in real timing systems is correlated according to a particular noise spectrum. 
Finally, we assume classical noise from the bath, whereas in reality quantum and other non-Markovian effects can lead to differences in the observed coherence decay.

While this work focuses on the NV center in diamond, the framework can be readily utilized to evaluate design parameters of integrated control signal sources for other solid-state spin qubits as well as other quantum platforms. 
All that is needed is an understanding of the intrinsic environmental noise spectrum of the quantum platform, so that it can be evaluated in conjunction with noise from the local oscillator. 
Similarly, this approach can be extended to evaluate design parameters for other quantum applications, such as quantifying the detection noise floor in quantum sensing \cite{Berzins2024}.
Depending on the application requirements, especially including the duration and number of pulses in the control sequence, quantitative bounds can be placed on the oscillator noise spectrum, including both its amplitude and frequency dependence, such that the system can be designed for optimal performance with minimal constraints.

\section{Acknowledgements}
We thank R. E. K. Fishman and J. Minnella for their assistance with photon correlation measurements.
This work was supported by the National Science Foundation (NSF) under award ECCS-1842655.
T.Y.H. acknowledges support from the Fontaine Society.
S.A.B. acknowledges support from an IBM PhD Fellowship.

\bibliography{references}

\clearpage

\section*{Supporting Information}
\subsection{Decoherence from nuclear spin bath and low-noise oscillator}
\begin{figure}[ht!]
    \renewcommand{\thefigure}{ S\arabic{figure}}
    \let\nobreakspace\relax
    \setcounter{figure}{0}
    \includegraphics[width=\linewidth]{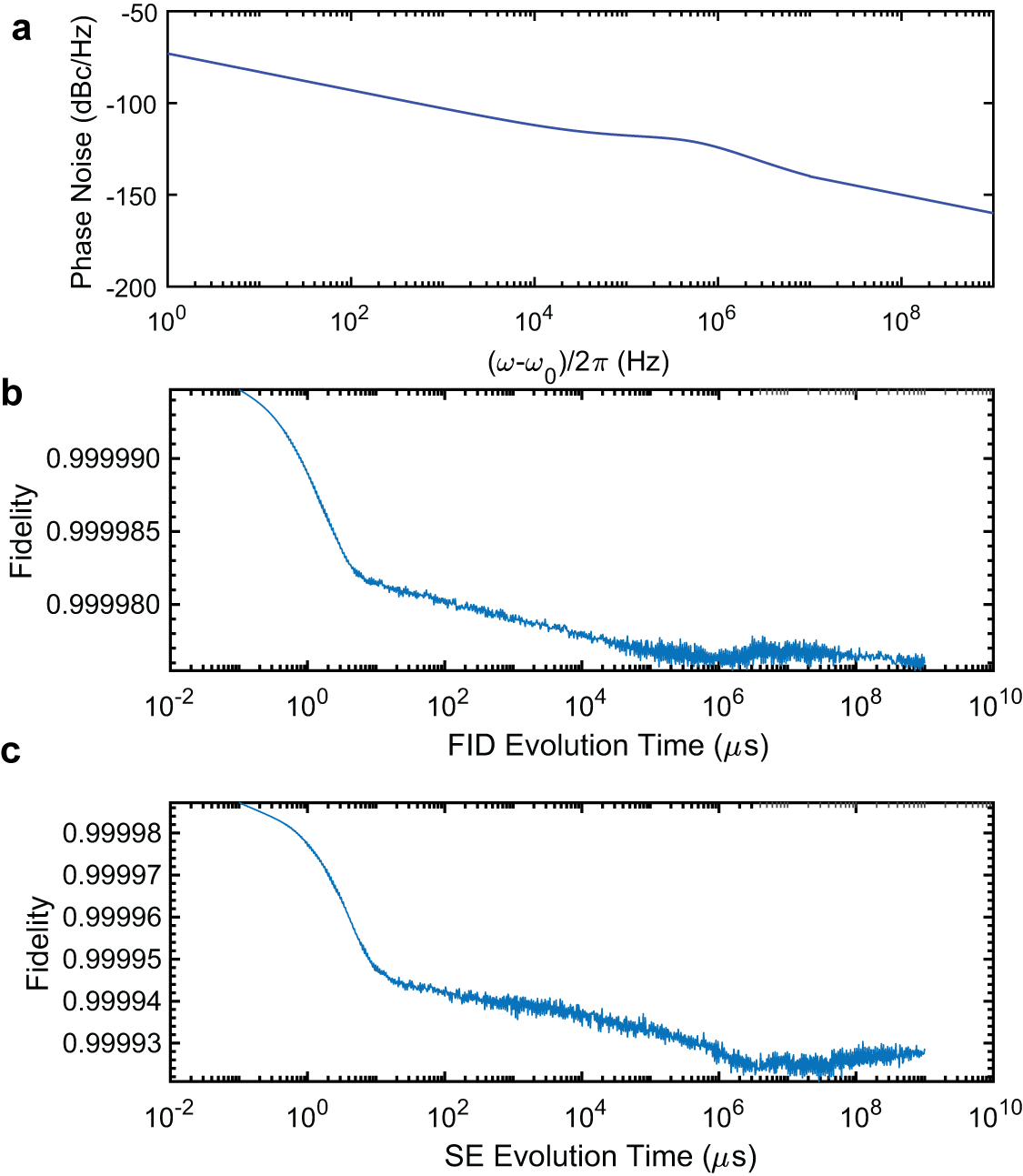}
    \caption[Central Spin Decoherence Induced by Low-noise Oscillator]{\label{fig:srsdeco}\textbf{Central Spin Decoherence Induced by Low-noise Oscillator.} \textbf{a}, phase spectrum reported by the manufacturer at \SI{1}{\giga\hertz}. \textbf{b-c}, Decoherence envelopes calculated from phase spectrum in (\textbf{a}) for (\textbf{b}) free-induction decay and (\textbf{c}) spin-echo.}
\end{figure}

For calculating $S_{z,\mathrm{electron}}$, we used $\Delta = \SI{6}{\mega\hertz}$ and $\tau_c = \SI{0.17}{\micro\second}$. For calculating $S_{z,\mathrm{nuclear}}$, we used $\Delta = \SI{6.2}{\mega\hertz}$ and $\tau_c = \SI{10}{\milli\second}$.

The reported phase spectrum specified by manufacturer for the low-noise oscillator (Stanford Research Systems, SG384) is shown in Fig.~\ref{fig:srsdeco}\textbf{a}. The corresponding decoherence envelopes calculated for free-induction decay (Fig.~\ref{fig:srsdeco}\textbf{b}) and spin-echo (Fig.~\ref{fig:srsdeco}\textbf{c}) indicate that there should be negligible contribution from the low-noise local oscillator towards decoherence of the central spin.

\begin{figure}[t!]
    \renewcommand{\thefigure}{ S\arabic{figure}}
    \let\nobreakspace\relax
    \includegraphics[width=\linewidth]{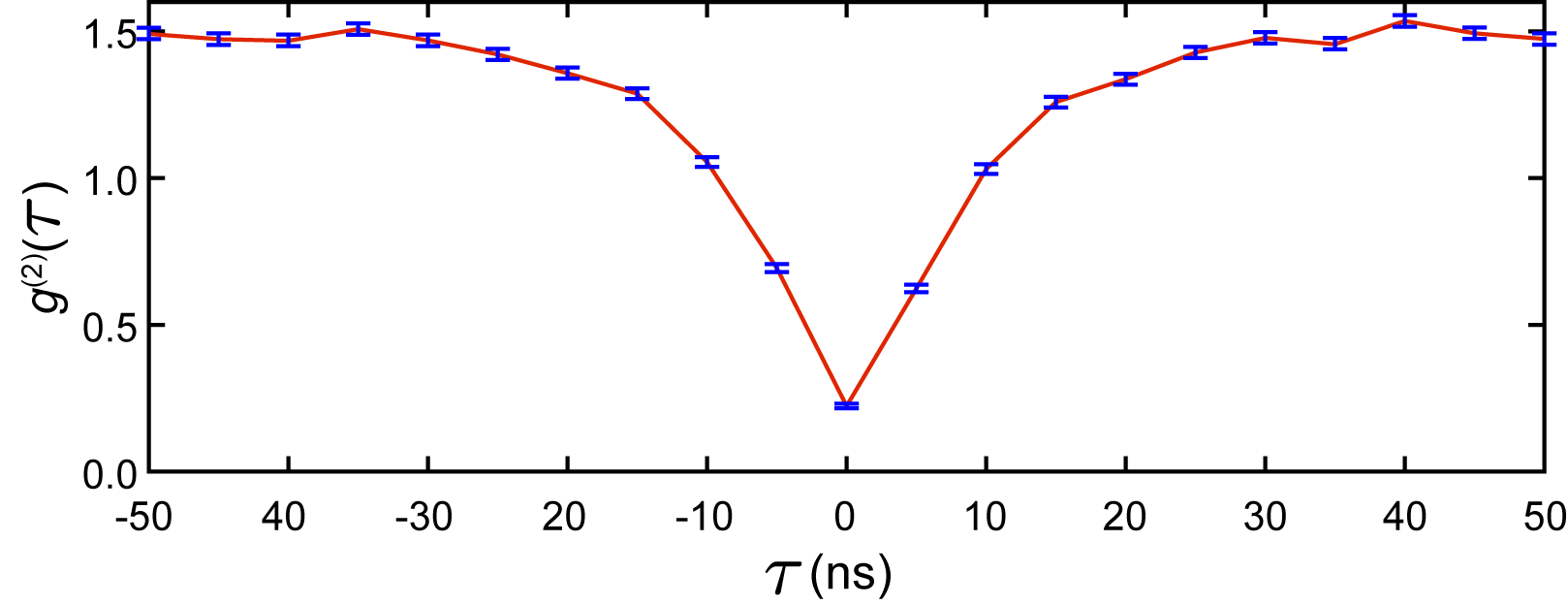}
    \caption[Second-order intensity correlation measurement]{\label{fig:g2}\textbf{Second-order intensity correlation measurement.} Measured (blue) and fitted (red) second-order intensity correlation measurement of the NV center studied in this work. Error bars represent the Poisson uncertainty in each bin of the correlation function.}
\end{figure}

\begin{figure}[t!]
    \renewcommand{\thefigure}{ S\arabic{figure}}
    \let\nobreakspace\relax
    \includegraphics[width=\linewidth]{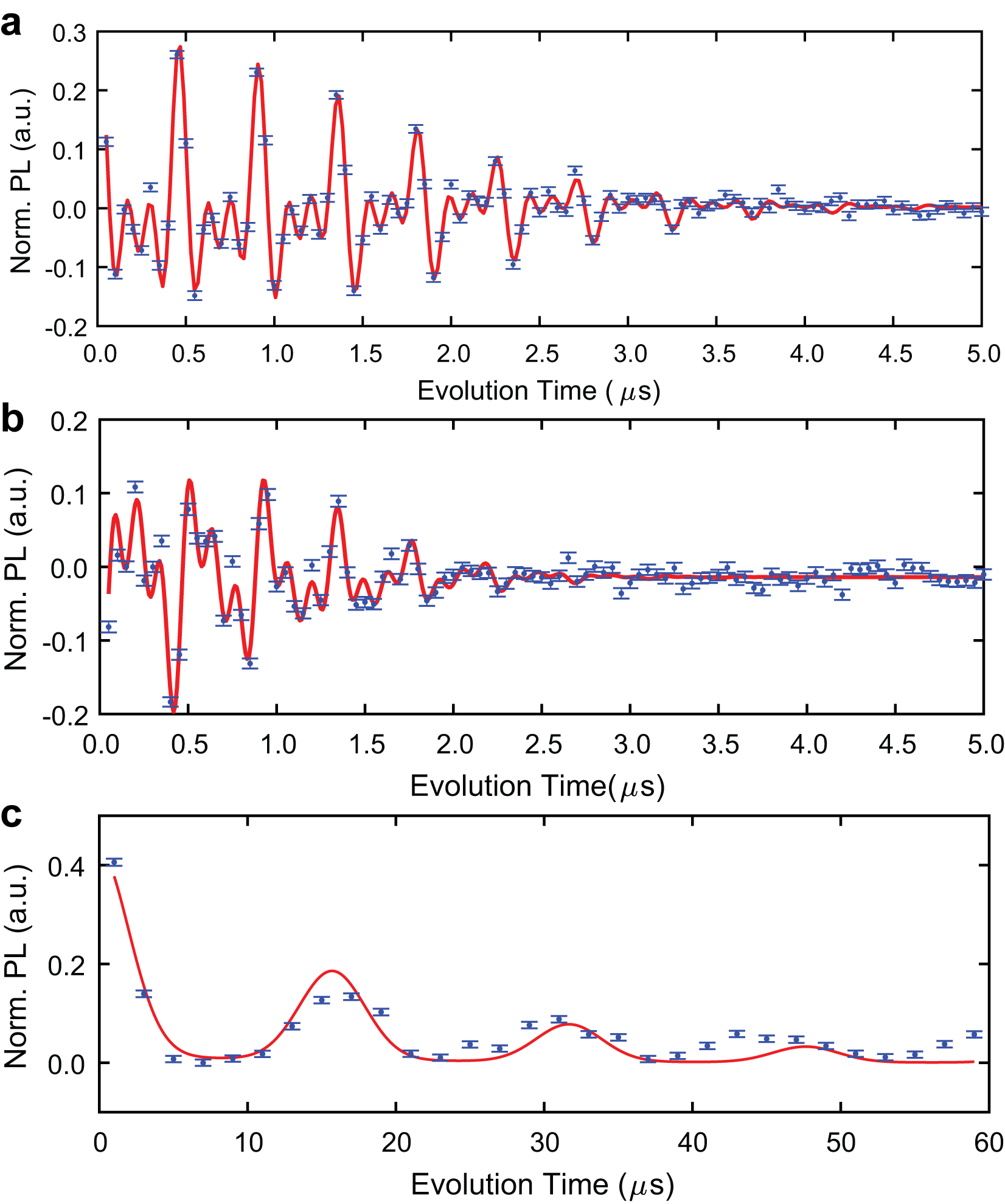}
    \caption[Fitting free-induction and spin-echo decays. ]{\label{fig:fits}\textbf{Fitting free-induction and spin-echo decays.} \textbf{a}, Measured (blue) and fitted (red) free-induction decay using the SG384. \textbf{b}, Measured (blue) and fitted (red) free-induction decay using the integrated VCO and 0 degree applied phase noise. \textbf{c}, Measured (blue) and fitted (red) spin-echo decay using the integrated VCO and 0 degree applied phase noise.}
\end{figure}

\subsection{NV center single-spin measurements}
The second-order intensity correlation measurement is shown in Fig.~\ref{fig:g2}. The measurement is fitted to a three-level empirical model and background-corrected according to the methods described in \cite{Fishman2023}. This measurement has not been corrected with the instrument response function.

Free-induction decay and spin-echo measurements were carried out on a custom-built scanning confocal microscope \cite{Hopper2020}. A coninuous-wave \SI{532}{\nano\meter} laser (Laser Quantum, Gem 532) was used as the excitation source and steered by a fast-steering mirror (Optics In Motion, OIM101). Emitted photons were filtered by a dichroic filter (Semrock, BrightLine FF560-FDi01) and counted by a single-photon avalanche diode (Laser Components, COUNT-T100 SPCM). The optical pulses were modulated by an artificial waveform generator (Tektronix, AWG520) while the control signal was modulated by another (Tektronix, AWG7102).

\begin{figure}[t!]
    \renewcommand{\thefigure}{ S\arabic{figure}}
    \let\nobreakspace\relax
    \includegraphics[width=\linewidth]{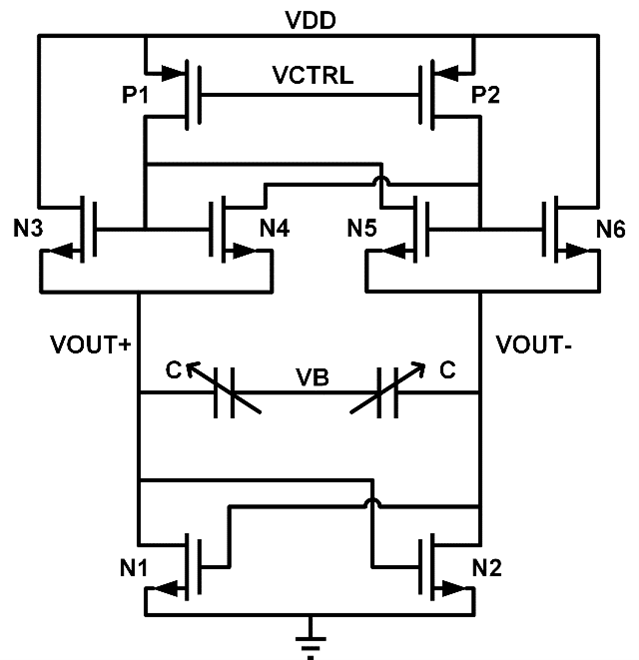}
    \caption[Schematic Diagram of the Voltage-controlled Oscillator]{\label{fig:vco}\textbf{Schematic Diagram of the Voltage-controlled Oscillator.} }
\end{figure}
Free-induction oscillations measured using the low-noise oscillator are shown in Fig.~\ref{fig:fits}\textbf{a}. Experimental values of $T_2^\ast$ are extracted from a fit to the free-induction decay data using the empirical model \cite{Hopper2020}:
\begin{equation}
\renewcommand{\theequation}{ S\arabic{equation}}
\setcounter{equation}{1}
\label{eqn:ramseyfit}
    I_{\mathrm{PL}} = C + A * e^{-(\tau/T^\ast_2)^2} \sum_{k=-1}^{1} \mathrm{cos}(2\pi(\delta-kA_{||})+\phi),
\end{equation}
\noindent where $I_{\mathrm{PL}}$ is the number of counted photons, $C$ is the dephased intensity level, $A$ is the signal amplitude, $\delta$ is the detuning from NV electron spin resonance, $A_{||} = \SI{2.16}{\mega\hertz}$ is the parallel hyperfine coupling to the $^{14}$N nuclear spin, and $\phi$ is the phase offset in radians. The spin-echo envelopes and corresponding fit measured using the low-noise oscillator were reported in Hopper \textit{et al}. (\cite{Hopper2020})

Fig.~\ref{fig:fits}\textbf{b-c} shows the measured and fitted free-induction and spin-echo decays using the integrated VCO with 0 degree applied phase noise.
Experimental values of $T_2$ are extracted from a fit to data using the empirical model consisting of a series of Gaussian revivals modulated by a decay envelope \cite{Shields2015}:
\begin{equation}
\renewcommand{\theequation}{ S\arabic{equation}}
\setcounter{equation}{2}
\label{eqn:hahnfit}
    I_{\mathrm{PL}} = C + A * e^{-(\tau/T_2)^n} \sum_{j=0}^{m} e^{-((\tau-jT_\mathrm{rev})/T_\mathrm{W})^2}.
\end{equation}
Here, $n$ is a free parameter for the stretched exponential, $m$ is the number of revivals measured, $T_\mathrm{rev}$ is the revival period, and $T_\mathrm{W}$ is the Gaussian width. The fitted $T^\ast_2$ from Fig.~\ref{fig:fits}(\textbf{b}) is $1.55 \pm \SI{0.23}{\micro\second}$. The fitted $T_2$ from Fig.~\ref{fig:fits}(\textbf{c}) is $18.81 \pm \SI{4.60}{\micro\second}$.

\begin{figure}[t!]
    \renewcommand{\thefigure}{ S\arabic{figure}}
    \let\nobreakspace\relax
    \includegraphics[width=\linewidth]{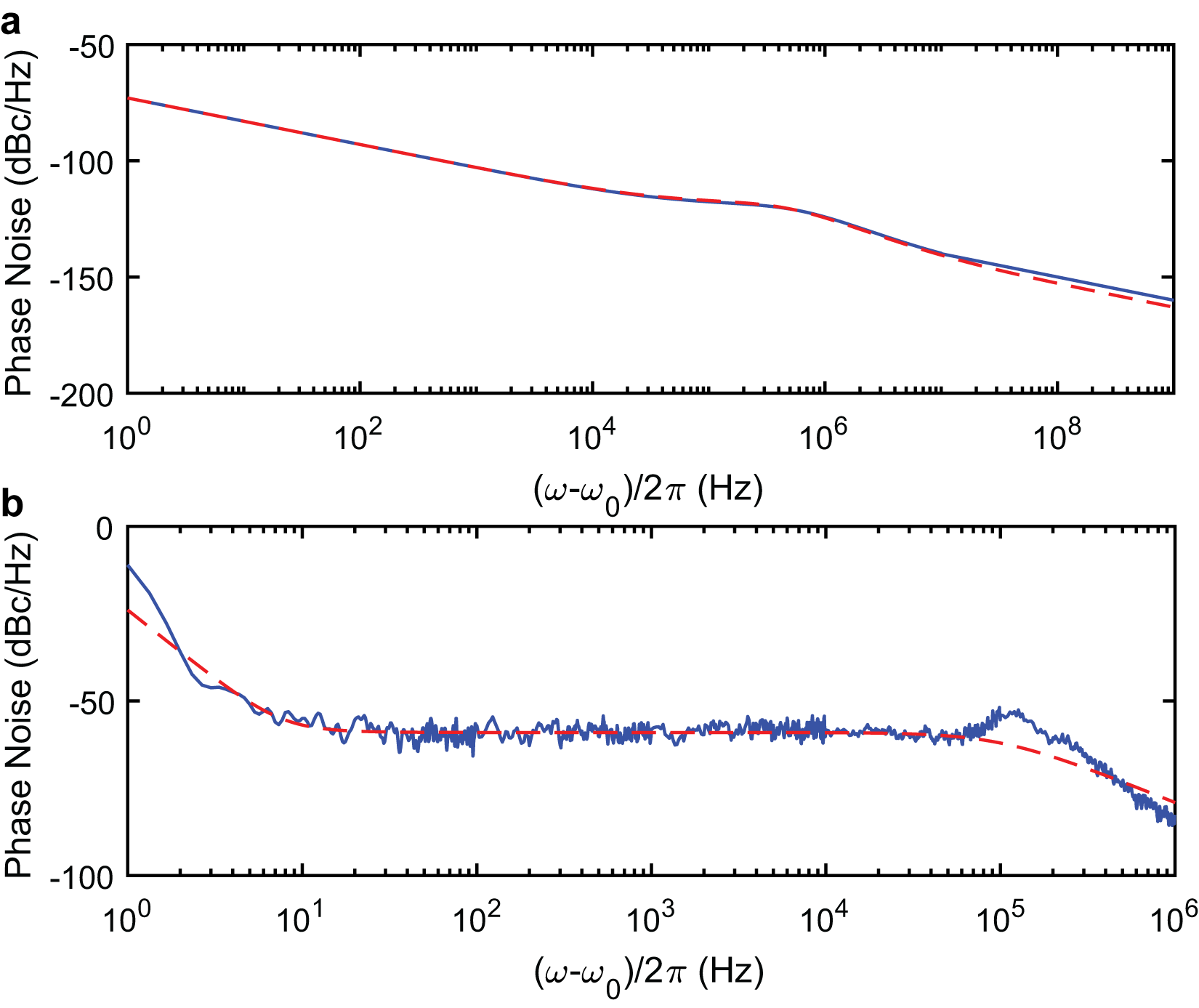}
    \caption[Modeled phase spectra]{\label{fig:modeledpn}\textbf{Modeled phase spectra.} \textbf{a-b}, Measured (blue) and modeled (red) single-sideband phase spectra of (\textbf{a}) the benchtop low-noise oscillator and (\textbf{b}) the integrated VCO.}
\end{figure}

\begin{figure*}[t!]
    \renewcommand{\thefigure}{ S\arabic{figure}}
    \let\nobreakspace\relax
    \includegraphics[width=\textwidth]{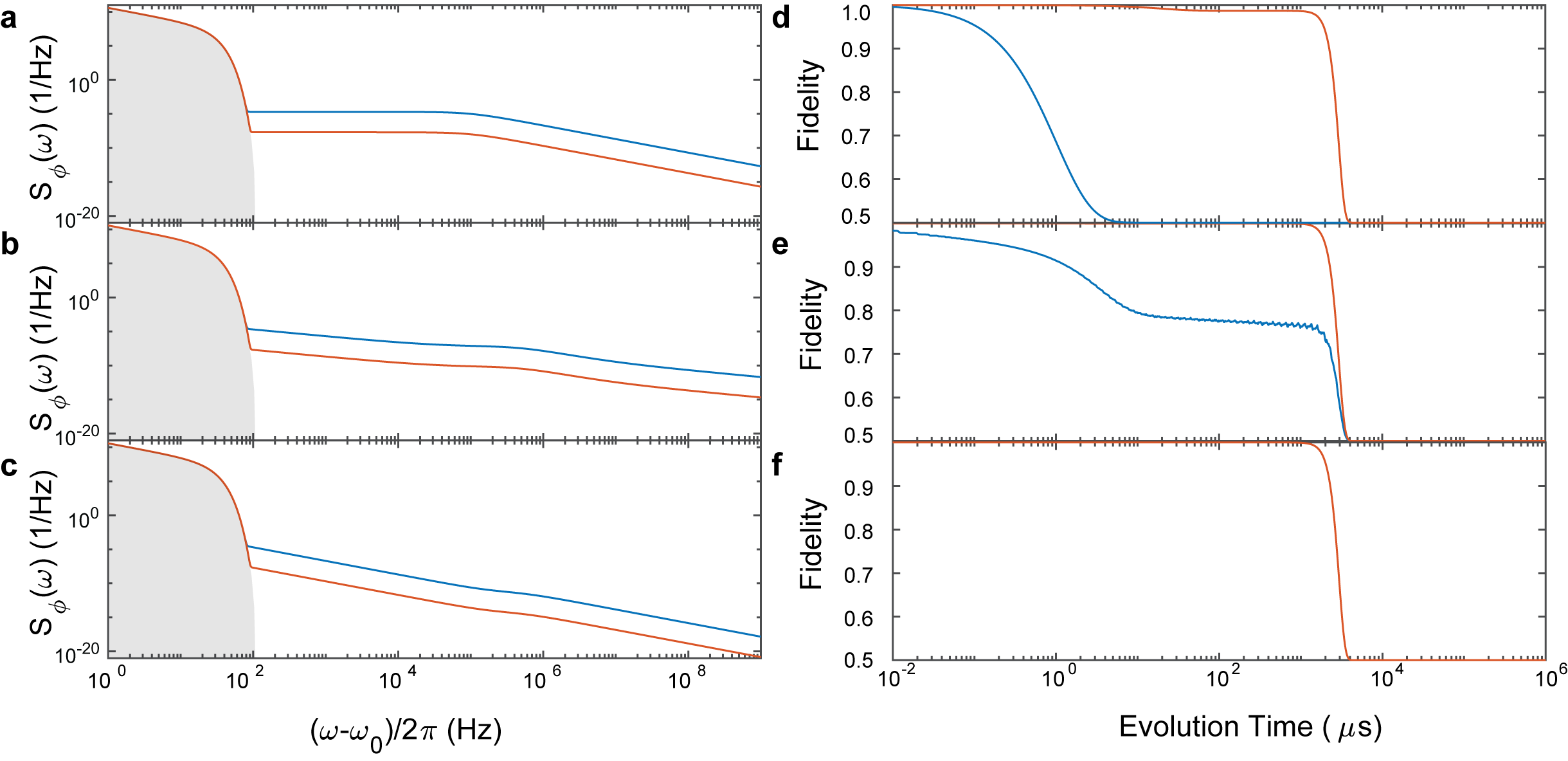}
    \caption{\label{fig:modeledwbath}\textbf{Calculated fidelity from modeled phase spectra of varying noise amplitude and frequency dependence.} \textbf{a-c}, Example phase spectra modeled after frequency dependencies of (\textbf{a}), the integrated VCO, (\textbf{b}), the benchtop synthesizer, and (\textbf{c}), $1/\omega^2$. The blue (red) traces have a phase amplitude of -50 (-80) dBc/Hz at $f = \SI{100}{\hertz}$.  Shaded regions denote noise spectrum of $^{13}$C nuclear spin bath. \textbf{d-f}, Calculated fidelity from CPMG sequences of $N = 8$ using the corresponding phase spectra modeled after frequency dependencies of (\textbf{d}), the integrated VCO, (\textbf{e}), the benchtop synthesizer, and (\textbf{f}), $1/\omega^2$. The traces in (\textbf{f}) overlap.}
\end{figure*}
\subsection{Voltage-controlled oscillator}

The schematic diagram of the voltage-controlled oscillator is presented on Fig.~\ref{fig:vco}. The architecture is based on cross-coupled transistor pair (N1-N2) and tunable active inductor (N3-N6, P1-P2) and varactors (C) \cite{Shirazi2010}. In this architecture, the active inductor is responsible for coarse frequency tuning (by adjusting node VCTRL) and varactors are responsible for fine frequency tuning (by adjusting node VB). The output of the system is differential (VOUT+ and VOUT-), where one output is connected to the spectrum analyzer for monitoring purpose and the other output is connected to the NV center. By injecting noise to VCTRL node, the phase noise of the VCO can be varied. The VCO is fabricated on 180-nm CMOS process for a 1.8V supply voltage.

\subsection{Modeling phase spectra}

\begin{table}[H]
\renewcommand{\thetable}{ S\arabic{table}}
\let\nobreakspace\relax
\centering
\begin{tabular}{l|c|c|c|c|c|c|c}
  & $h_{-2}$ & $h_{-1}$ & $h_{0}$ & $h_{1}$ & $h_{2}$ & $h_L$ & $\omega_L$\\
\hline
SG384 & 0 & 0 & 0 & 1e-25 & 0 & 3e-12 & 5e5\\
\hline
VCO & 2e-21 & 0 & 0 & 0 & 0 & 2.5e-6 & 1e5\\
\end{tabular}
\caption{\label{tab:coeffs} \textbf{Coefficients for modeling phase spectra.}}
\end{table}

Coefficients used to model the phase spectra of the SG384 and integrated VCO are shown in Table ~\ref{tab:coeffs}. Fig.~\ref{fig:modeledpn} shows the reported spectrum of the SG384 and the measured spectrum of the integrated VCO along with their respective modeled spectrum.

The phase spectrum of the integrated VCO was measured with the SG384 as the reference to its PLL. The spectrum is acquired using the same methods as the spectra presented in Fig.~5\textbf{a} of the main text, but with a resolution of \SI{1}{\hertz}.

To generate phase spectra of different starting noise levels, a scaling factor was applied to the modeled phase spectrum:

\begin{equation}
    \renewcommand{\theequation}{S\arabic{equation}}
    \setcounter{equation}{3}
    \tilde{S}_n(\omega) = \alpha_n S(\omega),
    \label{eqn:scalepn}
\end{equation}

\noindent where $n = [1..100]$ and $\alpha_n$ is spaced logarithmically. The resulting $\tilde{S}_n(\omega)$ are used to generate the calculations shown in Figs.~6 of the main text. 
Representative $\tilde{S}_n(\omega)$ curves modeled with frequency dependencies of the integrated VCO, benchtop frequency synthesizer, and $1/\omega^2$ are shown in conjunction with the nuclear spin bath in Fig.~\ref{fig:modeledwbath}.
The example amplitudes chosen are -50 and -80 dBc/Hz, and $N = 8$ for the calculations of fidelity. 

\subsection{CPMG filter function with timing jitter}
computing $|y_\mathrm{n-jitter}(\omega\tilde{\tau})|^2$ from Eqn.~(15) of the main text with a Gaussian distribution for $\delta_j$ gives,

\begin{widetext}
\begin{equation}
    \renewcommand{\theequation}{S\arabic{equation}}
    \setcounter{equation}{4}
    \label{jitterintegral}
    \int_{-\infty}^\infty \int_{-\infty}^\infty\cdots\int_{-\infty}^\infty\left| e^{-i\omega\delta_0} + (-1)^{N+1}e^{i\omega\tilde{\tau}_{N+1}} + 2\sum_{j=1}^{N}(-1)^je^{i\omega\tilde{\tau_j}} \right|^2 P(\delta_{N+1}) P(\delta_N) \ldots P(\delta_0) d\delta_0 \ldots d\delta_N d\delta_{N+1},
\end{equation}
\end{widetext}

\noindent where $P(\delta_j)$ is the Gaussian distribution of timing offsets. This expression can then be simplified to give Eqn.~(16) in the main text. 

\end{document}